\documentstyle[preprint,aps,epsf]{revtex}

\def\singlespace {\smallskipamount=3.75pt plus1pt minus1pt
                  \medskipamount=7.5pt plus2pt minus2pt
                  \bigskipamount=15pt plus4pt minus4pt
                  \normalbaselineskip=12pt plus0pt minus0pt
                  \normallineskip=1pt
                  \normallineskiplimit=0pt
                  \jot=3.75pt
                  {\def\smallskip {\vskip\smallskipamount}}
                  {\def\medskip   {\vskip\medskipamount}}
                  {\def\bigskip   {\vskip\bigskipamount}}
                  {\setbox\strutbox=\hbox{\vrule
                    height10.5pt depth4.5pt width 0pt}}
                  \parskip 7.5pt
                  \normalbaselines}
\def\middlespace {\smallskipamount=5.625pt plus1.5pt minus1.5pt
                  \medskipamount=11.25pt plus3pt minus3pt
                  \bigskipamount=22.5pt plus6pt minus6pt
                  \normalbaselineskip=22.5pt plus0pt minus0pt
                  \normallineskip=1pt
                  \normallineskiplimit=0pt
                  \jot=5.625pt
                  {\def\smallskip {\vskip\smallskipamount}}
                  {\def\medskip   {\vskip\medskipamount}}
                  {\def\bigskip   {\vskip\bigskipamount}}
                  {\setbox\strutbox=\hbox{\vrule
                    height15.75pt depth6.75pt width 0pt}}
                  \parskip 11.25pt
                  \normalbaselines}
\def\doublespace {\smallskipamount=7.5pt plus2pt minus2pt
                  \medskipamount=15pt plus4pt minus4pt
                  \bigskipamount=30pt plus8pt minus8pt
                  \normalbaselineskip=30pt plus0pt minus0pt
                  \normallineskip=2pt
                  \normallineskiplimit=0pt
                  \jot=7.5pt
                  {\def\smallskip {\vskip\smallskipamount}}
                  {\def\medskip   {\vskip\medskipamount}}
                  {\def\bigskip   {\vskip\bigskipamount}}
                  {\setbox\strutbox=\hbox{\vrule
                    height21.0pt depth9.0pt width 0pt}}
                  \parskip 15.0pt
                  \normalbaselines}

\tightenlines
\begin{document}
\preprint{
%\font\fortssbx=cmssbx10 scaled \magstep2
%\hbox to \hsize{
%\special{psfile=iulogo.ps
% hscale=8000 vscale=8000
% hoffset=-12 voffset=-2}
%\hskip.5in \raise.1in\hbox{\fortssbx University of Wisconsin - Madison}
\hfill$\vcenter{\hbox{\bf IUHET-422} \hbox{March
             2000}}$  }

\title{\vspace*{.75in}
Quark and Lepton Masses from a {\boldmath $U(1)\times Z_2$} Flavor Symmetry}

\author{Micheal S. Berger
\footnote{Electronic address:
berger@gluon.physics.indiana.edu}
and Kim Siyeon
\footnote{Electronic address:
siyekim@indiana.edu}}

\address{
Physics Department, Indiana University, Bloomington, IN 47405, USA}

\maketitle

\thispagestyle{empty}

\begin{abstract}
We show that solutions for the masses and mixings of the quarks and leptons 
based on a $U(1)\times Z_2$ horizontal symmetry are possible. The seesaw 
mechanism is shown to work consistently in the presence of the discrete
symmetry. The discrete symmetry results in the phenomenologically useful 
suppressions of elements of the Yukawa matrices. The quark and lepton masses, 
the CKM mixing angles, and the neutrino mixing angles are accommodated at the 
order of magnitude level.
\end{abstract}

\newcommand{\be}{\begin{equation}}
\newcommand{\ee}{\end{equation}}
\newcommand{\bea}{\begin{eqnarray}}
\newcommand{\eea}{\end{eqnarray}}

\newpage

\section{Introduction}
Experimental data on the quark and lepton sector masses and mixings may
provide a clue to the nature of new physics beyond the Standard Model (SM).
Masses and mixings are experimentally accessible, but as far as the SM is 
concerned, these parameters can be adjusted at will without 
destroying the consistency of the theory. 
Therefore any relationships between them must come from theoretical ideas
beyond those already contained in the SM, and the experimental data can 
guide us in narrowing down the choices and freedom in these ideas.

The recent data on the mixing of neutrinos has rekindled interest in models 
of fermion masses and mixings since it supplements the existing data from
the quark and charged lepton sectors. The neutrino observations have some
intriguing features that one might hope to explain. First of all, 
the neutrinos are very light in comparison to the other fermions. This 
suggests that a heavy mass scale may be involved that is providing a small
dimensionless number that is responsible for the small neutrino masses.
Secondly,
the atmospheric neutrino data\cite{superk} indicates that there is a large
mixing angle involved. This is in contrast to the small mixing 
(Cabibbo-Kobayashi-Maskawa of CKM) angles of the quark sector. Since in grand
unified theories (GUTs) the quarks and leptons are unified in representations
of the larger gauge theory, this dichotomy of small CKM angles with large 
mixing in the lepton sector provides hints as to how the fermion
masses might arise.
In fact the quark and charged lepton sectors show large hierarchies of masses.
This seems to indicate that there
might be a flavor symmetry whose spontaneous breaking might result in the 
generation of naturally small contributions resulting in the hierarchical
pattern of masses. One hope is that such a flavor symmetry can be implemented
to understand the masses and mixings detailed above as well as the 
long-standing evidence for solar neutrino oscillations.

In this paper we show solutions  to the quark and lepton masses and mixings
based on a $U(1)\times Z_2$ Abelian flavor symmetry are possible. 
A particular solution (with nontrivial $Z_2$ charges) we detail is entirely 
consistent with all the phenomenological constraints, and one can implement 
the seesaw mechanism to explain the light neutrino masses. 

The paper is organized as follows. In Section II we briefly review the 
approach of supersymmetric Abelian flavor (or horizontal) 
symmetries, and present 
the phenomenological requirements that must be satisfied in both the lepton
and quark sectors. In Section III we discuss how a discrete symmetry can be
used to overcome the constraints implied by a $U(1)$ symmetry, and show how
the discrete symmetry can suppress an entry to the extent that it has no 
impact on the leading order predictions for the masses and mixing angles.
In Sections IV and V we review how the light neutrino mass matrix is
independent of the horizontal charges of the singlet neutrinos for the case 
where the symmetry is $U(1)$. In Section V we also generalize the derivation
of the light neutrino mass matrix for the case where the horizontal symmetry
is $U(1)\times Z_2$. In Section VI we present a solution for the quark sector
that satisfies all the phenomenological requirements. Finally in Section VII
we present our conclusions.

\section{Flavor Symmetries}

The hierarchical structure of the fermion mass matrices strongly suggests
that there is a spontaneously broken family symmetry responsible for the 
suppression of Yukawa couplings.
In this paper we employ supersymmetric Abelian horizontal symmetries.
These flavor symmetries allow the 
fermion mass and mixing hierarchies to be naturally 
generated from nonrenormalizable 
terms in the effective low-energy theory.

The idea is quite simple and easily implemented\cite{fn}. 
There is some field $\Phi$
which is charged under a $U(1)$ family symmetry, and without loss of 
generality, we can assume that its charge is -1. There are terms contributing 
to effective Yukawa couplings for the quarks,
\bea
&&Q_i\overline{d}_jH_d\left ({{<\Phi>}\over {\Lambda _L}}\right )^{m_{ij}}
+Q_i\overline{u}_jH_u\left ({{<\Phi>}\over {\Lambda _L}}\right )^{n_{ij}}\;,
\eea
and the integer exponents $m_{ij}$ and $n_{ij}$ 
are easily calculated in terms of the 
horizontal symmetry charges of the quark and Higgs fields. For example, if 
we choose to have the Higgs fields to be uncharged, then the exponent 
$m_{ij}$ is just the sum of the horizontal charge of the fields 
$Q_i$ and $\overline{d}_j$. 
The hierarchy is generated from terms 
in the superpotential that carry integer 
charges $m_{ij},n_{ij}\geq 0$. If we call the small breaking parameter 
$\lambda$, then the generated terms for say the down quark 
Yukawa matrix will be of order $\lambda ^{m_{ij}}$.
The holomorphy of the superpotential forbids terms from 
arising with $m_{ij},n_{ij}<0$. 
A nice analysis of the possible approaches to explaining
the neutrino masses and mixings using $U(1)$ symmetries only is given in 
Ref.~\cite{lr}.
In the remainder of this section we outline the experimental constraints
that a solution employing the above idea must satisfy. 

\subsection{Phenomenological requirements for leptons}

The first phenomenological constraints we consider
involve the charged leptons whose
masses are the most precisely measured parameters of the quark-lepton sector.
For the experimental values for the masses, we require that 
\bea
{{m_\mu}\over {m_\tau}}\sim\lambda^2,
\ \ \ {{m_e}\over {m_\mu}}\sim\lambda^3\;, \label{chleptons}
\eea
where the small parameter is identified as the Cabibbo angle, i.e. 
$\lambda \sim 0.22$.

The remaining constraints on leptons involve the neutrino masses and mixings.
The most interesting aspect of the neutrino data is that the atmospheric 
neutrino mixing appears to be large, perhaps even maximal. It is then hard 
to understand a hierarchical pattern for the neutrino masses, since large
mixing should result when the neutrino masses are of roughly the same order 
of magnitude. The Super-Kamiokande data\cite{superk} suggest that 
\bea
&&\Delta m^2_{23}\sim 2.2\times 10^{-3}~{\rm eV}^2\;, \quad
\sin ^2 2\theta_{23}^\nu \sim 1\;, \label{atmos}
\eea
where the subscripts indicate the generations of neutrinos involved in the 
mixing (here we assume the mixing is between $\nu_\mu$ and $\nu_\tau$).

The solar neutrino flux can be explained by one of three distinct solutions.
Two of these involve matter-enhanced oscillation (MSW), while the third
involves vacuum oscillations (VO). The two MSW solutions are differentiated by
the size of the mixing angle, so one is usually called the small mixing angle
(SMA) solution, and the other is called the large mixing angle (LMA) solution.
The values required for the mixing parameters in each of these three cases
are shown in the table below.
\bea
\matrix{&\Delta m_{1x}^2\ [eV^2]&\sin^22\theta_{1x}\nonumber \\ \nonumber
{\rm MSW(SMA)}&5\times10^{-6}&6\times10^{-3}\\ \nonumber
{\rm MSW(LMA)}&2\times10^{-5}&0.8\\ \nonumber
{\rm VO}&8\times10^{-11}&0.8}
\eea

For example, consider the small mixing angle (SMA) solution of the 
solar neutrino problem. Combining the solar neutrino data with the atmospheric 
neutrino data, one requires then the following 
\bea
{{\Delta m^2_{1x}\over\Delta m^2_{23}}\sim\lambda^4,
\ \ \ \sin \theta_{12}^\nu \sim\lambda^2,
\ \ \ \sin \theta_{23}^\nu \sim \lambda^0,}
\eea
when the small parameter is taken to be the Cabibbo angle.
As explained by Grossman, Nir, Shadmi\cite{gns} and Tanimoto\cite{tan}, 
one can accommodate the phenomenological constraints on the neutrino masses 
and mixings as well as the charged
lepton masses
by postulating that there is a $U(1)\times Z_2$ horizontal symmetry.
We show how such a solution can be obtained in the seesaw mechanism in 
Section V.

\subsection{Phenomenological requirements for quarks}

Again taking the expansion parameter to be the 
Cabibbo angle, $\lambda=|V_{us}|$, 
then the experimental constraints\cite{pdg}
\bea
&&|V_{us}|=0.2196\pm 0.0023\;, \ \ |V_{cb}|=0.0395\pm 0.0017\;, \ \ 
\left |{{V_{ub}}\over {V_{cb}}}\right |=0.08\pm 0.02 \;,
\label{pdgdata}
\eea
on the CKM matrix can be identified in terms of powers of $\lambda$ by 
the following
\bea
&&|V_{us}|\sim \lambda \;, \ \ |V_{cb}|\sim \lambda ^2 \;, \ \ 
|V_{ub}| \sim
\lambda ^3 -\lambda ^4\;, \ \ 
\left |{{V_{ub}}\over {V_{cb}}}\right |\sim \lambda -\lambda^2 \;.
\label{ckmelem}
\eea
The constraint on $|V_{ub}/V_{cb}|$ can be expressed 
in a stronger way at 90\% confidence level as $0.25\lambda -0.5\lambda$.
One also has a constraint on 
the CKM elements from $B_d^0-\overline{B}_d^0$ mixing\cite{pdg},
\bea
&&|V_{tb}^*V_{td}|=0.0084\pm 0.0018\;, \label{pdgdata2}
\eea
which implies that 
\bea
&&|V_{td}|\sim \lambda ^3\;.\label{Vtd}
\label{ckmelem2}
\eea
It has been argued that $|V_{ub}|$ is more accurately 
given as $\sim \lambda ^4$ and taking it to be $\sim \lambda ^3$ (as we will
do) requires an unnatural cancellation. However, in our opinion, requiring
$|V_{ub}|\sim \lambda ^4$ is too restrictive for two reasons. Firstly, the 
fine-tuning required is much less if one uses an expansion parameter $\lambda $
somewhat less than 0.22, say 0.18. Secondly, since there are four parameters 
in the CKM matrix we are trying to predict, it is not unnatural that one of 
these would appear mildly fine-tuned, given $\lambda \sim 1/5$.

One can appreciate the nature of 
the cancellation in terms of the 
unitarity of the CKM matrix. This constraint requires
\bea
&&V_{ud}V_{ub}^*+V_{cd}V_{cb}^*+V_{td}V_{tb}^*=0\;,\label{unit}
\eea
so to leading order in $\lambda$ one has the relation\footnote{We use the 
notation $a\simeq b$ to indicate that $a$ and $b$ are equal to leading order
in the small parameter $\lambda$, while we use $a\sim b$ to indicate that 
$a$ and $b$ are the same order in $\lambda$.}
\bea
&&V_{ub}^*+V_{td}+V_{cd}V_{cb}^*\simeq 0\;.\label{tri}
\eea
Since $|V_{cd}|\sim \lambda$, $|V_{cb}|\sim \lambda^2$ and 
$|V_{td}|\sim \lambda ^3$, unitarity implies that to leading order 
one might expect $|V_{ub}|\sim \lambda ^3$ whereas the experimental data yields
a value somewhat suppressed\footnote{Equation~(\ref{tri}) 
represents the familiar unitarity triangle, and the cancellation can be 
reinterpreted in terms of the size of the CP asymmetry angle $\beta$.
Moreover, the unitarity triangle makes it clear how to interpret the 
three mixing angles and one CP-violating phase of the CKM matrix 
in terms of the four CKM elements in Eq.~(\ref{tri}). The amount of 
CP-violation is proportional to the size of the triangle.}.
One can show\cite{lns} 
that with $U(1)$ or $Z_2$ as components of the horizontal 
symmetry, one can suppress $|V_{ub}|$ (or $|V_{td}|$) 
relative to $\lambda ^3$ only by even powers of $\lambda$, 
so $|V_{ub}|\sim \lambda^4$ is not possible\footnote{In fact, the first 
(unsuccessful) solution in Section VI predicts $|V_{td}|\sim \lambda ^5$ and
the relation $|V_{us}|\simeq |V_{ub}/V_{cb}|$.}.
The additional suppression
one might desire to attribute to an additional power of $\lambda$ must in
fact be resulting from a mild cancellation.

There is a universal 
scaling factor associated with the renormalization group evolution
of the CKM angles $|V_{ub}|$ and $|V_{cb}|$ from the high (grand unified) scale
to the electroweak scale\cite{bbo}, but this scaling is not enough to change
the expectations for the relevant exponents of $\lambda $.
The mass ratios should satisfy
\bea
&&{{m_c}\over {m_t}}\sim \lambda ^4\;, \ \
{{m_u}\over {m_c}}\sim \lambda ^4\;, \ \
{{m_s}\over {m_b}}\sim \lambda ^2\;, \ \
{{m_d}\over {m_s}}\sim \lambda ^2\;. \label{quarkpheno}
\eea

To compare the predictions of flavor symmetries to these phenomenological 
constraints, one has to relate the CKM elements to the entries in the 
Yukawa matrices. The Yukawa matrices ${\bf U}$ and ${\bf D}$ can be 
diagonalized by biunitary
transformations
\begin{eqnarray}
{\bf U^{diag}}&=&V_u^L{\bf U}V_u^{R\dagger} \;, \\
{\bf D^{diag}}&=&V_d^L{\bf D}V_d^{R\dagger} \;.
\end{eqnarray}
The CKM matrix is then given by 
\begin{equation}
V\equiv V_u^LV_d^{L\dagger } \;.
\end{equation}
The left-handed
transformation matrices $V_u^L$ and $V_d^L$ can be defined in 
terms of three successive rotations in the (2,3), (1,3) and (1,2) sectors. 
These rotation angles of the transformation matrices can be expressed in 
terms of the elements of the Yukawa matrices as follows\cite{lns,hr}
%\begin{mathletters}
\bea
s_{12}^u&=&{{u_{12}}\over {\tilde{u}_{22}}}+{{u_{11}u_{21}^*}\over 
{|\tilde{u}_{22}^*|^2}}-{{u_{13}(u_{32}+u_{23}^*u_{22})}\over {\tilde{u}_{22}}}
-{{u_{11}u_{31}^*(u_{23}^*+u_{32}u_{22}^*)}\over {|\tilde{u}_{22}^*|^2}}\;, \\
s_{13}^u&=&u_{13}+u_{11}u_{31}^*+u_{12}(u_{32}^*+u_{22}^*u_{23})+u_{11}u_{21}^*
(u_{23}+u_{22}u_{32}^*)\;, \\
s_{23}^u&=&u_{23}+u_{22}u_{32}^*\;,\label{angles}
\eea
%\end{mathletters}
where $u_{ij}$ is the $i,j$th component of the up quark Yukawa matrix, 
${\bf U}/({\bf U})_{33}$,
and $\tilde{u}_{22}=u_{22}u_{33}-u_{23}u_{32}$.
There are corresponding expressions for the $s_{ij}^d$ in terms of the 
components of the down quark Yukawa matrix, ${\bf D}$ (which are slightly 
more complicated due to the fact that the (2,3) sector mixing in $V_d^R$ 
might be of
order one).
Clearly contributions to the CKM matrix elements can come from a number
of terms. We shall be interested in what follows in determining the leading
order contribution(s) to the CKM angles and the fermion masses.

\section{Texture Zeros}

The procedure of adopting a $U(1)$ horizontal symmetry 
introduces nontrivial relationships between the parameters in the 
quark and lepton mass matrices. This results because previously 
undetermined entries in the matrices are described in terms of a few 
parameters. For example, the quark (up and down) mass matrices are 
described by nine parameters, namely the $U(1)$ charges of the 
fields $Q_L$, $\overline{u}_R$, and $\overline{d}_R$. 
Relationships between the masses and mixing
angles are then obtained.

The motivation for including texture zeros in mass matrices was to derive
more relationships between the masses and mixings. The earliest of these was
the relationship between the Cabibbo angle and the first and second generation
down quark masses, $V_{us}\simeq \sqrt{m_d/m_s}$. The texture zeros 
responsible for this relation can be 
obtained in models where there is an additional discrete symmetry
that forbids their occurrence, for example.
Furthermore, these relationships might also include Clebsch-Gordon factors 
that allow one to obtain phenomenologically acceptable relationships: one of 
the earliest and most successful of these was the Georgi-Jarlskog 
model\cite{gj}, which was shown later\cite{dhr} to be successful in 
the case of electroweak scale supersymmetry (with the experimental
data available at that time). 
The guiding principle for the case where the Yukawa matrices are 
symmetric is this: 
the mass hierarchy 
is of order $\lambda ^4$ in the up quark matrices, and 
is of order $\lambda ^2$ in the down quark matrices 
(c.f. Eq.~(\ref{quarkpheno})). 
So the dominant contribution to the Cabibbo matrix should come from the 
down quark matrices (If the down quark matrix is symmetric and the 
1-1 component is suppressed, then one has the relation 
$|V_{us}|\simeq \sqrt{m_d/m_s}$), 
while the dominant contribution
to $|V_{cb}|\sim \lambda ^2$ should come from the diagonalization of 
the up quark matrices.
Furthermore the relation $|V_{ub}/V_{cb}|\simeq \sqrt{m_u/m_c}$
follow from suitable texture zero patterns\cite{hr}).

When the theory is supersymmetric, one can obtain zero entries 
in the mass matrices that are
sometimes called holomorphic or supersymmetric zeros. They arise
because the superpotential must be holomorphic, so entries that would get a 
contribution from $\Phi ^\dagger$ are absent. In terms of the mass matrices,
this simply means that there are no entries with the small parameter
$\lambda $ raised to a negative power. Rather entries, for which
the quantum numbers
would seem to require a negative exponent, are simply zero.

In this paper we want to introduce another concept that we will call a 
texture zero by flavor suppression. 
The horizontal symmetry we are 
considering here does not by itself give us any information on the order one
coefficients of the entries in the mass matrices.
When certain entries are suppressed because of the discrete symmetry, 
it can result that the entry is sufficiently suppressed that it does not
affect the leading order of the phenomenology. Equivalently this entry could
be replaced with an exact zero, 
and the leading order expectation for the masses and 
mixing angles would remain the same. Consider a simple $2\times 2$ example 
of quark mass matrices where there is a horizontal $U(1)$ symmetry, and
where the phenomenological constraints (listed below) 
are motivated by the (2,3) sector of the quark mass 
matrices. We require that  the mixing angle be $\sim \lambda^2$ and the
quark mass ratios satisfy
$m_{c}/m_{t}\sim \lambda^4$ and $m_{s}/m_{b}\sim \lambda^2$.
This can be obtained by assuming the charges $Q_L: 2,0$,
$\overline{u}_R: 2,0$, $\overline{d}_R: 0,0$:
\bea
&&{\bf U}\sim \pmatrix{\lambda ^4 & \lambda ^2 \cr
                   \lambda ^2 & \lambda ^0 }\;,\qquad
  {\bf D}\sim \pmatrix{\lambda ^2 & \lambda ^2 \cr
                   \lambda ^0 & \lambda ^0 }\;. \label{example}
\eea
This is a unique solution of $U(1)$ charges, and thereby determines already
relationships between the mixings and masses of the first generation.
The procedure for determining the exponents of $\lambda$ in a model with 
a $U(1)$ solution, clearly leads to the relations between exponents,
\bea
&&n_{ii}+n_{jj}=n_{ij}+n_{ji}\;,\label{simplerel}
\eea
for all $i,j=1,2,3$. 

Turning now to the case of a $U(1)\times Z_2$ symmetry, suppose the charges are
$Q_L: (2,1),(0,1)$, $\overline{u}_R: (1,0),(0,1)$, 
$\overline{d}_R: (0,1),(0,1)$,
and assume a common symmetry breaking parameter $\lambda$. We still 
achieve the hierarchies for ${\bf U}$ and ${\bf D}$ 
shown in Eq.~(\ref{example}).
So there are additional charge 
assignments that can be made that satisfy the 
phenomenological constraints.

However there is something more that adding a discrete symmetry can do.
Notice that in Eq.~(\ref{example}), 
the mixing $V_{cb}$ arises from contributions
from diagonalizing ${\bf U}$ and from diagonalizing ${\bf D}$, 
since both of these contributions
are of order $\lambda ^2$. We can however find stronger relationships 
if we can 
suppress one of these contributions. For example, if the 
mixing contribution from the (2,3) block of the 
up quark matrix is suppressed and the $({\bf D})_{22}$ entry is suppressed, 
then one has that the mixing angle is the same power of $\lambda$ as the 
square root of the mass ratio
$|V_{cb}|\sim \sqrt{m_c/m_t}$. (If the up quark matrix is known to be 
exactly symmetric, then one even knows that the order one coefficient of the 
leading contributions in $\lambda$ is the same, 
$|V_{cb}|\simeq \sqrt{m_c/m_t}$.)
Just this kind of suppression of the element
$({\bf U})_{23}$ can be obtained by employing a discrete symmetry. So if the 
exponent of $\lambda $ in $({\bf U})_{23}$ is sufficiently 
large that it plays no
role in determining the phenomenological predictions (to leading order), then
we say it is a texture zero. Returning to our example, take
the $U(1)\times Z_2$ charges to be
$Q_L: (3,1),(0,0)$, $\overline{u}_R: (1,1),(0,0)$, 
$\overline{d}_R: (1,0),(0,1)$.
Then one obtains the Yukawa matrices
\bea
&&{\bf U}\sim \pmatrix{\lambda ^4 & \lambda ^4 \cr
                   \lambda ^2 & \lambda ^0 }\;,\qquad
  {\bf D}\sim \pmatrix{\lambda ^5 & \lambda ^3 \cr
                   \lambda ^1 & \lambda ^1 }\;, \label{example2}
\eea
for which the phenomenological predictions in terms of powers of $\lambda $
are the same, but the mixing comes at leading order from diagonalizing
${\bf D}$.
We see that the relations, Eq.~(\ref{simplerel}), 
need not necessarily hold if one adds
a $Z_2$ symmetry.
Since it is the off-diagonal entries that are suppressed in ${\bf U}$, we can
define the texture pattern in the following schematic way,
\bea
&&{\bf U}\sim \pmatrix{X & 0 \cr
                   0 & X }\;,\qquad
  {\bf D}\sim \pmatrix{0 & X \cr
                   X & X }\;, \label{texexample}
\eea
where $X$ denotes a phenomenologically relevant entry, and the $0$ entries
are suppressed sufficiently that the exponent that appears there is irrelevant.
Even though the entries in the first column of the up and down quark matrices 
are not suppressed by the discrete symmetry, we denote these as zeros
because they do not contribute at leading order to either the 
phenomenologically relevant (left-handed)
mixing angles or the quark masses. 
We leave the categorization of what patterns of texture zeros one can employ
in $3\times 3$ matrices to a future work\cite{bk}.
 
\section{Neutrino Masses}

Assume that the lepton fields have charges under a $U(1)$ family symmetry
\bea
&&\begin{array}{c@{\quad}c@{\quad}c@{\quad}c@{\quad}c@{\quad}c
@{\quad}c@{\quad}c@{\quad}c}
\overline{e}_{R1} & \overline{e}_{R2} & \overline{e}_{R3} 
& \ell_{L1} & \ell_{L2} & \ell_{L3} 
& \overline{\nu}_{R1}^c & \overline{\nu}_{R2}^c & \overline{\nu}_{R3}^c \\
           E_1 & E_2 & E_3 & L_1 & L_2 & L_3 & {\cal N} _1 
                    & {\cal N} _2 & {\cal N} _3    \end{array}\;.\label{qnassn}
\eea
We assume here that the quantum numbers satisfy the hierarchies
$E_1\geq E_2\geq E_3\geq 0$, $L_1\geq L_2\geq L_3\geq 0$, and 
${\cal N}_1\geq {\cal N}_2\geq {\cal N}_3\geq 0$.

In the lepton sector, the effective Yukawa couplings 
come from nonrenormalizable terms, giving
\bea
&&\ell_{Li}\overline{e}_{Rj}H_d
\lambda^{p_{ij}}
+{1\over M_R}\ell_{Li}\ell_{Lj}H_uH_u
\lambda^{q_{ij}}\;,
\eea
where $M_R$ is the relevant high mass scale at which the light
neutrino masses are generated by the 
effective (nonrenormalizable) operator in the second term.
There are two cases we can consider: 
(1) the mechanism that gives rise to the light neutrino masses 
generates all possible contributions to the nonrenormalizable terms 
$\ell_{Li}\ell_{Lj}H_uH_u$.
In this case the exponent $q_{ij}$ that makes
the second term an invariant under the horizontal symmetry (before symmetry 
breaking) is just $q_{ij}=L_i+L_j$. So the light neutrino mass matrix is 
simply
\bea
&&m_\nu \sim \pmatrix{\lambda ^{2L_1} & \lambda ^{L_1+L_2} 
                                             & \lambda ^{L_1+L_3} \cr
                      \lambda ^{L_1+L_2} & \lambda ^{2L_2} 
                                             & \lambda ^{L_2+L_3} \cr
                      \lambda  ^{L_1+L_3} & \lambda ^{L_2+L_3} 
                                             & \lambda ^{2L_3}}
{{v_2^2}\over {\Lambda_L}}\;, \label{lightnu}
\eea
where $v_2$ is the electroweak scale vev of $H_u$ (and $v_1$ is the 
vev of $H_d$).
(2) The horizontal symmetry can play a role in the generation of the 
second term in which case it is not necessarily the case that the exponent
$q_{ij}$ is given by a naive inspection of the charges $L_i$ and $L_j$. The
seesaw mechanism for the generation of the light neutrino masses is such an 
example, and we explore it further in the next section.
In particular we show that with the appropriate assignment of heavy neutrino 
charges, ${\cal N}_i$, one can enhance the $(m_\nu)_{33}$ entry of the light
neutrino mass matrix.
 
\section{Neutrino Seesaw}

The neutrino seesaw mechanism is a 
popular way to explain the lightness
of the observed neutrino masses. It follows naturally from the group theory 
structure of the Standard Model (SM). We have left-handed neutrino doublets in 
the SM we can supplement by adding a right-handed neutrino singlet. 
In fact we have three generations, so the resulting masses will involve 
mass matrices.
The neutrino doublets can pair up with the singlets to form a Dirac mass 
matrix, $m_D$,
coming from the breakdown of the electroweak symmetry. The neutrino singlets
can pair up with themselves to form a $3\times 3$ Majorana matrix, $M_N$.
This mass matrix is expected to be superheavy; it is not generated by the 
electroweak symmetry breaking.
The group theory dictates that the neutrino doublets cannot pair up with each
other. So the result is a $6\times 6$ mass matrix of the form
\bea 
&&\pmatrix{0 & m_D^T  \cr
           m_D & M_N}\;,
\eea
and upon diagonalization of the $3\times 3$ light neutrino mass matrix is 
given by the seesaw formula
\bea
m_\nu=m_D (M_N)^{-1} m_D^T \label{eqn1}\;. \label{seesaw}
\eea

In the rest of this section we will explore the implications of assuming the
mass matrix entries are governed by an Abelian horizontal symmetry. A simple
result emerges for the case where the symmetry is $U(1)$, and some
interesting enhancements (or suppressions) can occur when the symmetry 
is extended to $U(1)\times Z_2$ which have some phenomenological usefulness.

\subsection{\boldmath $U(1)$}

Given lepton doublet charges $L_i$ and right-handed neutrino charges 
${\cal N}_i$
one has the following pattern for the neutrino Dirac mass matrix 
\bea
&&m_D \sim \pmatrix{\lambda ^{L_1+{\cal N}_1} & \lambda ^{L_1+{\cal N}_2} 
                                             & \lambda ^{L_1+{\cal N}_3} \cr
                      \lambda ^{L_2+{\cal N}_1} & \lambda ^{L_2+{\cal N}_2} 
                                             & \lambda ^{L_2+{\cal N}_3} \cr
                      \lambda  ^{L_3+{\cal N}_1} & \lambda ^{L_3+{\cal N}_2} 
                                             & \lambda ^{L_3+{\cal N}_3}}v_2\;,
\eea
and the following pattern for the Majorana mass matrix
\bea
&&M_N \sim \pmatrix{\lambda ^{2{\cal N}_1} & \lambda ^{{\cal N}_1+{\cal N}_2} 
                                 & \lambda ^{{\cal N}_1+{\cal N}_3} \cr
                   \lambda ^{{\cal N}_1+{\cal N}_2} & \lambda ^{2{\cal N}_2} 
                               & \lambda ^{{\cal N}_2+{\cal N}_3} \cr
         \lambda  ^{{\cal N}_1+{\cal N}_3} & \lambda ^{{\cal N}_2+{\cal N}_3} 
                                       & \lambda ^{2{\cal N}_3}}\Lambda _L\;.
\eea
Then one obtains the same form for the light neutrino mass matrix via
the see-saw formula $m_\nu=m_D (M_N)^{-1} m_D^T$ that was presented in the 
previous section in Eq.~(\ref{lightnu}).

It is easy to see that if one wants to have a hierarchy in light neutrino 
masses $m_{\nu_\mu}/m_{\nu_\tau}\sim \lambda ^2$, and large mixing in the (2,3)
generation in the lepton sector, one cannot rely on a $U(1)$ symmetry alone.
From Eq.~(\ref{lightnu}) one sees that $L_2=L_3+1$ is required to give the 
proper mass ratio. This then immediately prevents the large mixing from arising
in the neutrino sector, because the mixing is necessarily of order $\lambda$. 
However we must still examine the diagonalization of the charged lepton matrix.
Let the $U(1)$ charges of the charged lepton singlets be $E_i$, then the 
charged lepton matrix is 
\bea
m_{\ell^\pm} \sim \pmatrix{\lambda ^{L_1+E_1} & \lambda ^{L_1+E_2} 
                                             & \lambda ^{L_1+E_3} \cr
                      \lambda ^{L_2+E_1} & \lambda ^{L_2+E_2} 
                                             & \lambda ^{L_2+E_3} \cr
                      \lambda  ^{L_3+E_1} & \lambda ^{L_3+E_2} 
                                             & \lambda ^{L_3+E_3}}v_1\;.
\eea
The relevant mixing comes from the right hand side of this matrix,
$\lambda ^{L_2+E_3}/\lambda ^{L_3+E_3}$.
So the mixing parameter here is also order $\lambda $, since $L_2=L_3+1$.
Therefore the atmospheric neutrino mixing is necessarily of the order
$\sin \theta_{23}^\nu \sim \lambda$ in contradiction to the order one 
mixing observed (c.f. Eq.~(\ref{atmos})). 

\subsection{\boldmath $U(1)\times Z_2$}

We consider next the changes that can occur when a discrete symmetry is 
utilized. This avenue has been exploited to understand the large mixing
in the atmospheric neutrino oscillations together with a hierarchy in the 
neutrino masses, $m_{\nu_\mu}/m_{\nu_\tau}<<1$\cite{tan,gns}. It can 
also lead to an enhancement in the production of a matter/antimatter symmetry
in the early universe\cite{b-lepto}, if the heavy neutrinos are assumed to be
decaying asymmetrically. 
In the rest of this section we explain in detail how to 
implement the discrete symmetry with the seesaw mechanism so that the 
$m_{\nu_\mu}/m_{\nu_\tau}\sim \lambda ^2$, and large mixing results.
 
Take the following $U(1)\times Z_2$ charges for the lepton 
fields
\bea
&&\begin{array}{c@{\quad}c@{\quad}c@{\quad}c@{\quad}c
@{\quad}c@{\quad}c@{\quad}c@{\quad}c}
\overline{e}_{R1} & \overline{e}_{R2} & \overline{e}_{R3} 
& \ell_{L1} & \ell_{L2} & \ell_{L3} 
& \overline{\nu}_{R1} & \overline{\nu}_{R2} & \overline{\nu}_{R3} \\
           (E_1,0) & (E_2,0) & (E_3,0) & (L_1,0) & (L_2,0) & (L_3-1,1) 
& ({\cal N}_1,0) & ({\cal N}_2,0) & ({\cal N}_3-1,1)    \end{array}\;.
\eea
Assume the symmetry breaking is 
characterized by the single expansion parameter $\lambda$.
The formulas given above for the heavy neutrino mass matrix, $M_N$, 
the neutrino Dirac mass matrix, $m_D$, and the resulting light neutrino mass 
matrix, $m_\nu$ are modified. With the above assignments one finds that 
\bea
&&M_N \sim \pmatrix{\lambda ^{2{\cal N}_1} & \lambda ^{{\cal N}_1+{\cal N}_2} 
                                & \lambda ^{{\cal N}_1+{\cal N}_3} \cr
                   \lambda ^{{\cal N}_1+{\cal N}_2} & \lambda ^{2{\cal N}_2} 
                                    & \lambda ^{{\cal N}_2+{\cal N}_3} \cr
          \lambda  ^{{\cal N}_1+{\cal N}_3} & \lambda ^{{\cal N}_2+{\cal N}_3} 
                                    & \lambda ^{2{\cal N}_3-2}}\Lambda _L\;,
\eea
so that 
\bea
&&(M_N)^{-1} \sim \pmatrix{\lambda ^{-2{\cal N}_1} & 
                  \lambda ^{-{\cal N}_1-{\cal N}_2} 
                              & \lambda ^{-{\cal N}_1-{\cal N}_3+2} \cr
              \lambda ^{-{\cal N}_1-{\cal N}_2} & \lambda ^{-2{\cal N}_2} 
                                   & \lambda ^{-{\cal N}_2-{\cal N}_3+2} \cr
   \lambda  ^{-{\cal N}_1-{\cal N}_3+2} & \lambda ^{-{\cal N}_2-{\cal N}_3+2} 
                               & \lambda ^{-2{\cal N}_3+2}}\Lambda _L^{-1}\;.
\eea
So the effect of the discrete symmetry in our case is to enhance the 3-3 entry 
of the $M_N$ matrix, and thereby alter the results for the third row and the 
third column on the inverse matrix, $(M_N)^{-1}$.
With our charge assignments, one also has an enhanced entry in the 3-3 
component of the neutrino Dirac mass matrix,
\bea
&&m_D \sim \pmatrix{\lambda ^{L_1+{\cal N}_1} & \lambda ^{L_1+{\cal N}_2} 
                                             & \lambda ^{L_1+{\cal N}_3} \cr
                      \lambda ^{L_2+{\cal N}_1} & \lambda ^{L_2+{\cal N}_2} 
                                             & \lambda ^{L_2+{\cal N}_3} \cr
                      \lambda  ^{L_3+{\cal N}_1} & \lambda ^{L_3+{\cal N}_2} 
                                    & \lambda ^{L_3+{\cal N}_3-2}}v_2\;,
\eea
It is easy to see then that the light neutrino mass matrix $m_\nu$
is modified so that only the 3-3 entry is enhanced as desired,
\bea
&&m_\nu \sim \pmatrix{\lambda ^{2L_1} & \lambda ^{L_1+L_2} 
                                             & \lambda ^{L_1+L_3} \cr
                      \lambda ^{L_1+L_2} & \lambda ^{2L_2} 
                                             & \lambda ^{L_2+L_3} \cr
                      \lambda  ^{L_1+L_3} & \lambda ^{L_2+L_3} 
                                             & \lambda ^{2L_3-2}}
{{v_2^2}\over {\Lambda_L}}\;. \label{lightnuenhance}
\eea
A phenomenologically viable solution has been presented in
Ref.~\cite{gns}: taking $L_1=3$, $L_2=L_3=1$, $E_1=5$, $E_2=4$, and $E_3=2$
yields mass matrices of the form
\bea
&&m_\nu \sim \pmatrix{\lambda ^6 & \lambda ^4 
                                             & \lambda ^4 \cr
                      \lambda ^4 & \lambda ^2 
                                             & \lambda ^2 \cr
                      \lambda  ^4 & \lambda ^2 
                                             & 1}
{{v_2^2}\over {\Lambda_L}}\;, 
\qquad
m_{\ell^\pm} \sim \pmatrix{\lambda ^8 & \lambda ^7 
                                             & \lambda ^5 \cr
                      \lambda ^6 & \lambda ^5 
                                             & \lambda ^3 \cr
                      \lambda  ^6 & \lambda ^5 
                                             & \lambda ^3}v_1\;,\label{lepsoln}
\eea
which give the correct orders of magnitude for the small mixing angle MSW 
solution 
\bea
{{\Delta m^2_{12}\over\Delta m^2_{23}}\sim\lambda^4,
\ \ \ \sin \theta _{12}^\nu \sim\lambda^2,\ \ \ 
\sin \theta_{23}^\nu \sim \lambda^0,}
\eea
and the correct orders of magnitude
for the charged lepton mass ratios, Eq.~(\ref{chleptons}).
It also gives $\sin \theta_{13}^\nu \sim\lambda^2$.
In fact it has been shown\cite{gns} that one can 
obtain a VO solution as well 
by a proper quantum number assignment to the lepton fields.
 
Without introducing more theoretical input (e.g. grand unified theories) to 
relate the quark and lepton charges, we cannot say more about which of the
solutions is the correct one. We have shown here that the lepton sector
phenomenology and the neutrino seesaw mechanism
is consistent with a $U(1)\times Z_2$
symmetry, and in fact a hierarchy
in the neutrino parameters $m_{\nu_\mu}<<m_{\nu_\tau}$ requires the extra 
$Z_2$. Furthermore we have shown in detail 
how to implement the neutrino mass enhancement in the seesaw mechanism.
In the next section, we show that the quark sector phenomenological constraints
also admit a solution consistent with Eqs.~(\ref{ckmelem}) 
and (\ref{quarkpheno}), 
and with a $U(1)\times Z_2$ symmetry.

\section{A {\boldmath $U(1)\times Z_2$} solution}

Our solution to the quark Yukawa matrices is based upon
the Elwood-Irges-Ramond (EIR) solution\cite{eir} obtained with
a $U(1)$ flavor symmetry. Here we show that one can impose a texture
pattern by choosing quark fields to be charged under the new $Z_2$ 
symmetry. EIR obtained the Yukawa matrices
\bea
&&{\bf U}\sim \pmatrix{\lambda ^8 & \lambda ^5 & \lambda ^3 \cr
                 \lambda ^7 & \lambda ^4 & \lambda ^2 \cr
                 \lambda ^5 & \lambda ^2 & \lambda ^0}\;,\qquad
  {\bf D}\sim \pmatrix{\lambda ^4 & \lambda ^3 & \lambda ^3 \cr
                 \lambda ^3 & \lambda ^2 & \lambda ^2 \cr
                 \lambda ^1 & \lambda ^0 & \lambda ^0}\;.\label{eirsoln}
\eea
The EIR solution was obtained by the $U(1)$ charges (after adding appropriate
overall constants to each field which don't affect the hierarchy pattern) 
\bea
\matrix{i=&1&2&3\nonumber \\ \nonumber
Q_L:&4&3&1\\ \nonumber
\overline{u}_R:&4&1&-1\\ \nonumber
\overline{d}_R:&0&-1&-1}\;.
\eea
The CKM elements can be expressed in terms of the Yukawa matrix elements,
%\begin{mathletters}
\bea
|V_{us}|&=&\left ({{d_{12}}\over {\tilde{d}_{22}}}
-{{d_{13}d_{32}}\over 
{\tilde{d}_{22}}}\right )
-\left ({{u_{12}}\over {\tilde{u}_{22}}}-{{u_{13}u_{32}}\over 
{\tilde{u}_{22}}}\right )\;, \label{Vusangles} \\
|V_{cb}|&=&d_{23}+d_{22}d_{32}^*-u_{23}\;, \label{Vcbangles} \\
|V_{ub}|&=&(d_{13}+d_{12}d_{32}^*-u_{13})-\left ({{u_{12}}\over 
{\tilde{u}_{22}}}-
{{u_{13}u_{32}}\over {\tilde{u}_{22}}}\right )(d_{23}+d_{22}d_{32}^*-u_{23})\;,
\label{Vubangles} \\
|V_{td}|&=&-(d_{13}+d_{12}d_{32}^*-u_{13})+\left ({{d_{12}}\over 
{\tilde{d}_{22}}}-
{{d_{13}d_{32}}\over {\tilde{d}_{22}}}\right )(d_{23}+d_{22}d_{32}^*-u_{23})\;,
\label{Vtdangles}
\eea
%\end{mathletters}
where it is understood that there are possible phases associated with each 
term on the right hand sides of the equations. From Eq.~(\ref{Vubangles}), 
one sees that $|V_{ub}|$ 
is receiving contributions in the EIR solution
of order $\lambda ^3$ from both $u_{13}$ and $d_{13}$, as well as from 
the final term 
\bea
&&\left ({{u_{12}}\over {\tilde{u}_{22}}}-
{{u_{13}u_{32}}\over {\tilde{u}_{22}}}\right )V_{cb}\;.\label{singleterm}
\eea
Then the experimental value for $|V_{ub}|$ must arise from a  
partial cancellation of these three contributions.  

Tanimoto showed that a solution involving a $U(1)\times Z_2$ symmetry is not
possible if the Cabibbo mixing, $|V_{us}|$, arises from the diagonalization 
of the down quark Yukawa matrix\cite{tan}.
The Cabibbo mixing in our scheme comes from diagonalizing the up quark 
matrix ${\bf U}$. 
Our first attempt
has the following assignments for the 
quantum numbers of 
the quark fields
\bea
\matrix{i=&1&2&3\nonumber \\ \nonumber
Q_L:&(4,0)&(2,1)&(0,1)\\ \nonumber
\overline{u}_R:&(4,0)&(1,0)&(0,1)\\ \nonumber
\overline{d}_R:&(0,0)&(0,1)&(0,1)}\;,
\eea
which is easily related to the EIR $U(1)$ assignment above by replacing the 
$Z_2$ charge with $+1$ for $Q_L$ and with $-1$ for 
$\overline{u}_R$ and $\overline{d}_R$. 
One can always substitute $0\leftrightarrow 1$ for $Z_2$ charges without
affecting the results.
We obtain the following Yukawa matrices for the up and down quarks
\bea
&&{\bf U}\sim \pmatrix{\lambda ^8 & \lambda ^5 & \lambda ^5 \cr
                 \lambda ^7 & \lambda ^4 & \lambda ^2 \cr
                 \lambda ^5 & \lambda ^2 & \lambda ^0}\;,\qquad
  {\bf D}\sim \pmatrix{\lambda ^4 & \lambda ^5 & \lambda ^5 \cr
                 \lambda ^3 & \lambda ^2 & \lambda ^2 \cr
                 \lambda ^1 & \lambda ^0 & \lambda ^0}\;.
\eea
The mass matrices can be obtained by multiplying these Yukawa matrices by the
relevant vev ($v_1$ for ${\bf D}$ and $v_2$ for ${\bf U}$). The 
intergenerational hierarchy, $m_b/m_t\sim \lambda ^3$, is then accounted for
either by $\tan \beta =v_2/v_1$, and/or by increasing the $U(1)$ charges for
$\overline{d}_R$.
It is straightforward to check that these matrices have the texture pattern
\bea
&&{\bf U}\sim \pmatrix{X & X & 0 \cr
                 X & X & X \cr
                 0 & X & X}\;,\qquad
  {\bf D}\sim \pmatrix{X & 0 & 0 \cr
                 0 & X & X \cr
                 0 & X & X}\;.\label{texture}
\eea
We refer to this type of suppression as 
a 3-texture zero solution, since three entries are suppressed by the charge
assignments in the discrete symmetry.
Referring back to Eq.~(\ref{Vubangles}), one can see that
the dominant contribution to $|V_{ub}|$ 
comes only from the third term and is of order $\lambda ^3$. 
This solution was motivated by the desire\footnote{Since the experimental 
data for 
$|V_{cb}|=a\lambda ^2$ already requires the order one coefficient to be less 
than one, $a\simeq 0.6$\cite{tan}, it is more likely that this final term
(which includes another order one coefficient we will call $b$) 
will give agreement with
the experimental data, $|V_{ub}|=ab\lambda ^3$ with $b$ somewhat smaller than
one. Furthermore it is then consistent
with the experimental data on $|V_{ub}/V_{cb}|=0.08\pm 0.02 =b\lambda $ for
$b\simeq 0.4$. So the combination of order one coefficients satisfy
the required relation, $ab\simeq \lambda$.} to derive that $|V_{ub}|$ is 
proportional to $|V_{cb}|$; this can be achieved 
if the first term in parentheses in
Eq.~(\ref{Vubangles}) is suppressed, and this requires a texture zero
in the (1,3) position of both ${\bf U}$ and ${\bf D}$. Consequently
one only has a contribution from the final term ($u_{13}\sim \lambda ^5$ 
and $d_{13}\sim \lambda ^5$). But then 
$|V_{td}|\sim \lambda ^5$ is inconsistent with Eq.~(\ref{Vtd}), 
and unitarity (Eq.~(\ref{tri})) requires the 
relation
\bea
&&|V_{us}|\simeq \left |{{V_{ub}}\over {V_{cb}}}\right |\;,\label{relation}
\eea
which is also not supported by the experimental data, Eq.~(\ref{pdgdata}). 
Clearly the 3-texture zero 
pattern in Eq.~(\ref{texture}) is too restrictive.

We can relax the problematic constraint, Eq.~(\ref{relation}), by removing
the texture zero in the (1,3) position of the up quark matrix.
Consider the following texture zero pattern
\bea
&&{\bf U}\sim \pmatrix{0 & X & X \cr
                 X & X & X \cr
                 X & X & X}\;,\qquad
  {\bf D}\sim \pmatrix{X & 0 & 0 \cr
                 0 & X & X \cr
                 0 & X & X}\;.\label{texture2}
\eea
from 
\bea
\matrix{i=&1&2&3\nonumber \\ \nonumber
Q_L:&(4,0)&(2,1)&(0,1)\\ \nonumber
\overline{u}_R:&(6,1)&(2,0)&(0,0)\\ \nonumber
\overline{d}_R:&(0,0)&(0,1)&(0,1)}
\eea
We obtain the following Yukawa matrices for the up and down quarks
\bea
&&{\bf U}\sim \pmatrix{\lambda ^{11} & \lambda ^6 & \lambda ^4 \cr
                 \lambda ^8 & \lambda ^5 & \lambda ^3 \cr
                 \lambda ^6 & \lambda ^3 & \lambda ^1}\;,\qquad
  {\bf D}\sim \pmatrix{\lambda ^4 & \lambda ^5 & \lambda ^5 \cr
                 \lambda ^3 & \lambda ^2 & \lambda ^2 \cr
                 \lambda ^1 & \lambda ^0 & \lambda ^0}\;.\label{quarksoln}
\eea
As is clear from the texture pattern in Eq.~(\ref{texture2}), the Cabibbo angle
is arising in the up quark matrix ${\bf U}$. However one avoids the 
uncomfortable relation $|V_{us}|\sim \sqrt{m_u/m_c}$ because the matrix is
not symmetric. All phenomenological constraints are satisfied by this solution
with $|V_{ub}|$ receiving contributions of order $\lambda ^3$ from only the 
$u_{13}$ term and the last term in Eq.~(\ref{Vubangles}).
The matrices in Eq.~(\ref{lepsoln}) and (\ref{quarksoln}) show that
nontrivial $Z_2$ charges can be assigned to the quark and lepton fields, and
that all phenomenological requirements can be met.

\section{Conclusion}
We have shown that one can explain all the masses and mixings of the quarks and
leptons with a $U(1)\times Z_2$ symmetry. The phenomenological requirements can
be met if the Cabibbo mixing in the two light generations is generated 
in the up quark mass matrix. This runs counter to the  bias of assuming 
that the Cabibbo mixing is coming from the down quark mass matrix so that the
relation $|V_{us}|\simeq \sqrt{m_d/m_s}$ is obtained. 
This prejudice should be abandoned in the context of these Abelian flavor 
symmetries, because the resulting Yukawa matrices need not be symmetric. 
and $m_u/m_c\sim \lambda ^4$ is a reasonable hierarchy even with the leading
contribution to the Cabibbo angle $|V_{us}|$ coming from the up quark matrix. 

The advantages of employing the $U(1)\times Z_2$ as a flavor symmetry are the 
following:

\begin{itemize}

\item One can understand a mass hierarchy 
$m_{\nu_\mu}/m_{\nu_\tau}\sim \lambda^2$ and large atmospheric neutrino mixing
$\sin \theta _{23}^\nu \sim 1$, without invoking unnatural cancellations.

\item The discrete symmetry can be implemented consistently with the 
neutrino seesaw mechanism to give the necessary neutrino mass hierarchy.

\item The source for CKM mixing angles in terms of the original parameters
in the Yukawa matrices is reduced via the presence of texture zeros.
For example $|V_{ub}|$ arises from a single contribution in 
Eq.~(\ref{Vubangles}), since the other contributions are suppressed by a 
flavor suppression. The EIR model has the Cabibbo angle, 
$|V_{us}|\sim \lambda $, arising at leading order from both the up and down 
quark matrices (c.f. Eqs.~(\ref{Vusangles}) and (\ref{eirsoln})). 
Our solution suppresses the 
contribution from the down quark Yukawa matrix, and thus the leading 
contribution arises entirely in the up quark matrix.

\item The discrete $Z_2$ symmetry can enhance the lepton 
asymmetry generated by the decay of heavy right-handed neutrinos\cite{b-lepto}.
This can be exploited to straightforwardly 
explain the baryon asymmetry of the Universe.

\end{itemize}

A discrete flavor symmetry offers some attractive features for generating
phenomenologically reasonable models. We leave to future work the question of
whether these models can be embedded in a larger theoretical structure.
 
\section*{Acknowledgments}

This work was supported in part by the U.S.
Department of Energy
under Grant No. 
No.~DE-FG02-91ER40661.

\end{document}